\documentclass[fleqn,usenatbib]{mnras}

\usepackage{newtxtext,newtxmath}

\usepackage[T1]{fontenc}

\DeclareRobustCommand{\VAN}[3]{#2}
\let\VANthebibliography\thebibliography
\def\thebibliography{\DeclareRobustCommand{\VAN}[3]{##3}\VANthebibliography}

\usepackage{graphicx}	
\usepackage{amsmath}	
\usepackage{romannum}
\usepackage{textcomp}
\usepackage{gensymb}
\usepackage{enumitem}
\usepackage{hyperref}
\usepackage{multirow}
\usepackage[flushleft]{threeparttable}
\usepackage{subcaption}

\def\Rn2 {\Romannum{2} }
\def\p {\textquotesingle}

\title[The afterglow optical spectral shape of GRB201015A]{Evolution of the afterglow optical spectral shape of GRB 201015A in the first hour: evidence for dust destruction}

\author[Komesh et al.]{Toktarkhan Komesh,$^{1,2}$\thanks{E-mail: toktarkhan.komesh@nu.edu.kz}
Bruce Grossan,$^{1,3}$
Zhanat Maksut,$^{1}$
Ernazar Abdikamalov,$^{1,4}$\newauthor
Maxim Krugov$^{5}$
and George F. Smoot$^{1,6,7,8,9,10}$
\\
$^{1}$Energetic Cosmos Laboratory, Nazarbayev University, Astana 010000, Kazakhstan\\
$^{2}$Faculty of Physics and Technology, Al-Farabi Kazakh National University, Almaty, 050040, Kazakhstan\\
$^{3}$Space Sciences Laboratory, University of California, Berkeley, CA 94720, USA\\
$^{4}$Department of Physics, Nazarbayev University, 53 Kabanbay Batyr ave, Astana 010000, Kazakhstan\\
$^{5}$Fesenkov Astrophysical Institute, Almaty, 050020, Kazakhstan\\
$^{6}$Institute for Advanced Study Hong Kong University of Science and Technology, Clear Water Bay, Kowloon, Hong Kong\\
$^{7}$Universit\'e Sorbonne Paris Cit\'e, Laboratoire APC-PCCP, Universit\'e Paris Diderot, {\it emeritus}\\
$^{8}$Department of Physics, University of California, California, USA, {\it emeritus}\\
$^{9}$Universit\'e de Paris, Laboratoire Astroparticule et Cosmologie, F-75013 Paris, France, {\it emeritus}\\
$^{10}$Donostia International Physics Center, University of the Basque Country UPV/EHU, E-48080 San Sebastian, Spain
}

\date{MNRAS:
Accepted 2023 February 15. Received 2023 February 14; in original form 2022 November 3}

\pubyear{2022}

\begin{document}
\label{firstpage}
\pagerange{\pageref{firstpage}--\pageref{lastpage}}
\maketitle

\begin{abstract}

Instruments such as the ROTSE, TORTORA, Pi of the Sky, MASTER-net, and others have recorded single-band optical flux measurements of gamma-ray bursts starting as early as $\thicksim$ 10 seconds after gamma-ray trigger. The earliest measurements of optical spectral shape have been made only much later, typically on hour time scales, never starting less than a minute after trigger, until now. 
     Beginning only 58 seconds after the \emph{Swift} BAT triggerred on GRB201015A, we observed a sharp rise in optical flux to a peak, followed by a power law temporal decay, $\propto t^{-0.81 \pm 0.03}$. Flux was measured simultaneously in three optical  bands, g\p, r\p, and i\p,  using our Burst Simultaneous Three-channel Imager (BSTI) on the NUTTelA-TAO telescope. 
Our data during the decay show strong colour evolution from red to blue, with a change in the optical log slope of $+0.72 \pm 0.14$; during this time the X-ray log slope remained constant. 
We did not find evidence for a two-component jet structure or a transition from reverse to forward shock 
or a prompt emission component
that would explain this change in slope.
We find that the majority of the optical spectral slope evolution is consistent with a monotonic decay of extinction, evidence of dust destruction. Assuming a constant source spectral slope and an SMC-like extinction curve, we derive a change in the local extinction $A_\mathrm{v}^\mathrm{local}$ from $\thicksim$0.8 mag to 0.3 mag in $\thicksim$2500 seconds. 
This work shows that significant information about the early emission phase is being missed without such early observations with simultaneous multi-band instruments.

\end{abstract}

\begin{keywords}
gamma-ray burst: individual: GRB 201015A -- dust, extinction
\end{keywords}



\section{Introduction}

The prompt phase of a gamma-ray burst (GRB), that during bright and rapidly-varying $\gamma$-ray emission, is directly associated with a relativistic jet
        \citep[e.g.,][]{2004RvMP...76.1143P,2009ARA&A..47..567G}
and typically lasts $\sim$ a minute for Long-type GRB 
    \citep{2011ApJS..195....2S}. 
Later emission, stemming from the interaction of the explosion with the surrounding medium, is called the afterglow \citep[e.g.,][]{1997Natur.386..686V}. 
 
The single-band optical flux of GRBs has been observed as early as $\sim 10$ seconds after the onset of $\gamma$-ray emission; multiple instruments, including ROTSE, TORTORA, Pi of the Sky, and MASTER-net \citep[e.g.,][]{2008Natur.455..183R, 2000ApJ...542..819K, 2009ApJ...702..489R} have made such very early single-band measurements of GRB optical emission.
Such observations, intended to capture the prompt phase emission, automatically respond to 
alerts from the space-based $\gamma$-ray Burst Alert Telescope \citep[BAT,][]{2005SSRv..120..143B} aboard the Neil Gehrels \emph{Swift} observatory. 
However, the earliest measurements of the spectral shape were typically made much later, on hour time scales, in the afterglow phase \citep[e.g.,][]{2020GCN.29033....1P, 2011A&A...534A.108K, 2011A&A...535A..57F}. While the afterglow is well studied on this time scale \citep[nominally simple forward shock emission; e.g.,][]{2021ApJ...918...12F, 2010MNRAS.402...46M}, no optical spectral shape measurements \citep{2019JHEAp..23...14G,2009MNRAS.398.1936S} have been made in the prompt phase, or in the transition between these phases.
Early light curve studies (without optical spectra) have suggested that different jet components, and/or combinations of reverse shock, may explain the emission \citep[e.g.,][]{2008Natur.455..183R, 2014Sci...343...38V}.
In addition, the photodestruction of dust in the environments of GRBs is expected to occur early on \cite[e.g.,][]{2001ApJ...563..597F,2003ApJ...585..775P}.

In order to make early optical spectral shape measurements of GRBs triggered by BAT or similar instruments,
(i) observations must be made with a very fast-pointing, automatically-triggered telescope (in order to begin measurements within $\sim$ 1 minute)
\footnote{Ultra-wide field telescopes covering a large fraction of the sky, without pointing, would eventually record GRBs; however, with a typical sensitivity $<$ 10 mag in 10 seconds \citep{2014PASP..126..885G}, until the sensitivity of these instruments is improved by many magnitudes,  they will miss the vast majority of prompt and early GRB emission.}, 
and (ii) measurements must be made simultaneously in multiple bands (because the variability of these transients is much faster than, e.g., filter wheel motion in this phase). We designed and built a unique instrument, the Burst Simultaneous Three-channel Imager \citep[BSTI,][]{2020SPIE11447E..9IG}, mounted on the 700 mm aperture Nazarbayev University Transient Telescope at Assy-Turgen Astrophysical Observatory (NUTTelA-TAO), to make these measurements.  The system can point and track any celestial target above 15$^\circ$ altitude in $\leq$ 8 s, responding automatically to \emph{Swift} and other real-time GRB alerts, with time resolution down to $\sim 0.1$ seconds. 
We have been following up on GRB alerts since September 2020. We observed GRB 201015A starting only 58 seconds after
the BAT trigger, measuring in three Sloan filter bands, g', r', and i' \citep{2020GCN.28674....1G}.
Our early measurements, from the first minute to the first hour after the trigger, show changes in the emission processes not typically found later; these measurements offer diagnostics of both
the physical processes within the outflow and the environment of the progenitor \citep[e.g.,][]{2013A&A...557A..12Z, 2013ApJ...774...13L, 2008ApJ...686.1209M, 2007A&A...469L..13M, 2006Natur.442..172V}. In this work, our aim is to understand the early afterglow as distinct from the later phase.

GRB 201015A is of unusually low luminosity compared to most Long GRBs, with a prompt isotropic-equivalent energy of E$_\mathrm{iso}\thicksim10^{50}$ erg. 
However, it is consistent with the E$_\mathrm{p,i}$ vs. E$_\mathrm{iso}$ correlation (Amati relation) for long type bursts \cite[e.g.,][]{2020GCN.28668....1M,2022icrc.confE.797S}.
 
We report and discuss the NUTTelA-TAO
optical observations of GRB 201015A, together with reported X-$\gamma$ ray and other optical data.
        Based on terminology from, e.g. \cite{2014Sci...343...38V}, we define the early afterglow phase as the first hour after the BAT trigger, and the late afterglow thereafter.
        We concentrate on the  changes of temporal and spectral slopes in the early afterglow phase.
Interpretations of the data in terms of emission mechanisms and components are discussed in Section \ref{discussion}.
Throughout the paper we define $\alpha$ and $\beta$, the temporal and spectral log slopes, as $f_\nu \propto  t^{\alpha}\nu^{\beta}$.

\section{Observations and data} 
\label{observation}

We started observations of the field of GRB201015A at UT 2020 October 15, 22:51:11, 58 seconds after the BAT trigger, using the NUTTelA-TAO and BSTI.
The observational details are presented in Table \ref{tbl:2}. The cameras were operated with exposures of 7.5 seconds in the first 60 seconds, then exposures of 15 seconds until 1635 seconds, and exposures of 30 seconds thereafter.
The 5 sigma upper limit sensitivities are 17.7, 18.2 and 17.7 mag for most g\p, r\p \ and i\p \ filter images, respectively, in most 15 second exposure images.
The weather was clear during the observations.
Calibration was done with 5 bright Pan-STARRS catalog stars on our images. 
We tested our photometry by checking consistency of standard star magnitudes and colors with catalog values, and by checking our measurements of simulated stars (of precisely known flux) added to our  images.

We used the published i\p \ band data of the Binospec imager and spectrograph mounted on the MMT 6.5-meter telescope \citep[henceforth MMT,][]{2020GCN.28676....1R} on Mount Hopkins, Arizona, and r\p \ and g\p \ bands of the AZT-20 telescope of Assy-Turgen Astrophysical Observatory \citep[henceforth AZT-20,][]{2020GCN.28673....1B}.
 The BAT and XRT data  were obtained through the  website of the  UK \emph{Swift} Science Data Centre at the University of Leicester \footnote{\href{https://www.swift.ac.uk/burst_analyser/01000452/}{https://www.swift.ac.uk}}.

\begin{table*}
    \caption{NUTTelA-TAO observations of GRB201015A \label{tbl:2}}
    \centering
    \begin{tabular}{cccccccc}
    \hline
    \hline
	 T$_{mid}$ (s)    & t$_{exp}$ (s)  &	\multicolumn{3}{c}{Manitudes}   & \multicolumn{3}{c}{Flux (mJy)}     \\
		& 		& 		g\p (error)       & 	    r\p (error)         & 		i\p (error) & 	g\p (error)  & 	r\p (error)& i\p (error)                \\
	\hline
	88      &    60    &    18.47	(0.28)      &       17.98	(0.12)     &       17.84	(0.17)  &   0.49	(0.13)  &  0.54	(0.06) &   0.50	(0.08) \\
	143     &    15    &    17.95	(0.24)      &       17.44	(0.11)     &       16.89	(0.11)  &   0.80	(0.17)  &  0.88	(0.09) &   1.19	(0.12) \\
	158     &    15    &    17.62	(0.18)      &       17.02	(0.08)     &       16.79	(0.09)  &   1.09	(0.18)  &  1.31	(0.09) &   1.31	(0.11) \\
	173     &    15    &    17.72	(0.18)      &       17.30	(0.09)     &       16.72	(0.08)  &   0.99	(0.17)  &  1.01	(0.08) &   1.40	(0.11) \\
	188     &    15    &    17.55	(0.17)      &       16.92	(0.07)     &       16.60	(0.08)  &   1.15	(0.18)  &  1.44	(0.09) &   1.55	(0.11) \\
	203     &    15    &    17.47	(0.14)      &       17.08	(0.07)     &       16.73	(0.08)  &   1.24	(0.16)  &  1.23	(0.08) &   1.38	(0.10) \\
	218     &    15    &    17.38	(0.15)      &       16.80	(0.06)     &       16.32	(0.07)  &   1.35	(0.19)  &  1.60	(0.09) &   2.01	(0.12) \\
	233     &    15    &    17.63	(0.19)      &       17.10	(0.08)     &       16.63	(0.08)  &   1.07	(0.19)  &  1.22	(0.09) &   1.52	(0.11) \\
	248     &    15    &    17.31	(0.15)      &       16.84	(0.07)     &       16.45	(0.08)  &   1.43	(0.20)  &  1.55	(0.10) &   1.78	(0.12) \\
	263     &    15    &    17.71	(0.19)      &       17.08	(0.08)     &       16.53	(0.08)  &   1.00	(0.17)  &  1.23	(0.09) &   1.66	(0.12) \\
	278     &    15    &    17.57	(0.18)      &       17.02	(0.08)     &       16.56	(0.08)  &   1.14	(0.19)  &  1.31	(0.10) &   1.61	(0.12) \\
	293     &    15    &    17.64	(0.21)      &       17.16	(0.10)     &       16.50	(0.08)  &   1.07	(0.21)  &  1.15	(0.10) &   1.71	(0.13) \\
	308     &    15    &    17.60	(0.20)      &       16.94	(0.07)     &       16.67	(0.09)  &   1.10	(0.20)  &  1.41	(0.10) &   1.46	(0.12) \\
	323     &    15    &    17.50	(0.17)      &       17.18	(0.09)     &       16.55	(0.08)  &   1.21	(0.19)  &  1.13	(0.10) &   1.63	(0.12) \\
	338     &    15    &    17.34	(0.16)      &       16.95	(0.08)     &       16.63	(0.09)  &   1.40	(0.20)  &  1.40	(0.10) &   1.51	(0.13) \\
	353     &    15    &    17.51	(0.19)      &       17.04	(0.08)     &       16.62	(0.09)  &   1.20	(0.22)  &  1.28	(0.09) &   1.52	(0.12) \\
	368     &    15    &    17.95	(0.26)      &       16.94	(0.07)     &       16.44	(0.07)  &   0.80	(0.19)  &  1.40	(0.09) &   1.81	(0.11) \\
	383     &    15    &    17.51	(0.16)      &       17.25	(0.09)     &       16.80	(0.10)  &   1.20	(0.18)  &  1.06	(0.09) &   1.29	(0.11) \\
	398     &    15    &    17.64	(0.19)      &       17.03	(0.08)     &       16.52	(0.08)  &   1.06	(0.19)  &  1.30	(0.10) &   1.68	(0.12) \\
	413     &    15    &    17.63	(0.16)      &       17.39	(0.10)     &       16.75	(0.09)  &   1.07	(0.16)  &  0.93	(0.09) &   1.35	(0.11) \\
	428     &    15    &    17.81	(0.23)      &       17.04	(0.08)     &       16.73	(0.10)  &   0.91	(0.19)  &  1.28	(0.09) &   1.38	(0.12) \\
	443     &    15    &    17.77	(0.21)      &       17.29	(0.10)     &       16.77	(0.10)  &   0.94	(0.19)  &  1.02	(0.09) &   1.33	(0.12) \\
	480     &    60    &    17.53	(0.12)      &       17.23	(0.06)     &       16.95	(0.07)  &   1.18	(0.13)  &  1.07	(0.06) &   1.13	(0.07) \\
	540     &    60    &    17.88	(0.13)      &       17.43	(0.06)     &       17.08	(0.07)  &   0.85	(0.10)  &  0.90	(0.05) &   1.00	(0.06) \\
	600     &    60    &    17.96	(0.14)      &       17.61	(0.07)     &       17.33	(0.08)  &   0.80	(0.10)  &  0.76	(0.05) &   0.79	(0.06) \\
	660     &    60    &    18.43	(0.19)      &       17.80	(0.08)     &       17.41	(0.09)  &   0.52	(0.09)  &  0.63	(0.05) &   0.74	(0.06) \\
	758     &    135   &    18.38	(0.16)      &       18.07	(0.08)     &       17.48	(0.08)  &   0.54	(0.08)  &  0.50	(0.04) &   0.69	(0.05) \\
	893     &    135   &    18.43	(0.16)      &       18.13	(0.09)     &       17.84	(0.11)  &   0.52	(0.08)  &  0.47	(0.04) &   0.50	(0.05) \\
	1028    &    135   &    18.83	(0.22)      &       18.32	(0.10)     &       17.94	(0.11)  &   0.36	(0.07)  &  0.39	(0.04) &   0.45	(0.05) \\
	1163    &    135   &    18.61	(0.18)      &       18.43	(0.10)     &       18.21	(0.14)  &   0.43	(0.07)  &  0.36	(0.03) &   0.35	(0.04) \\
	1298    &    135   &    19.31	(0.32)      &       18.60	(0.12)     &       18.11	(0.12)  &   0.23	(0.07)  &  0.31	(0.03) &   0.39	(0.04) \\
	1433    &    135   &    19.01	(0.24)      &       18.69	(0.13)     &       18.24	(0.14)  &   0.30	(0.07)  &  0.28	(0.03) &   0.34	(0.05) \\
	1568    &    135   &    19.28	(0.32)      &       18.76	(0.14)     &       18.36	(0.15)  &   0.24	(0.07)  &  0.26	(0.03) &   0.31	(0.04) \\
	2100    &    270   &    19.29	(0.19)      &       18.92	(0.12)     &       18.52	(0.15)  &   0.23	(0.04)  &  0.23	(0.03) &   0.26	(0.04) \\
	2370    &    270   &    19.25	(0.23)      &       19.06	(0.16)     &       18.51	(0.18)  &   0.24	(0.05)  &  0.20	(0.03) &   0.27	(0.04) \\
	2640    &    270   &    19.25	(0.19)      &       19.29	(0.14)     &       18.77	(0.16)  &   0.24	(0.04)  &  0.16	(0.02) &   0.21	(0.03) \\
	2910    &    270   &    19.40	(0.22)      &       19.47	(0.19)     &       18.85	(0.21)  &   0.21	(0.04)  &  0.14	(0.02) &   0.20	(0.04) \\
	3180    &    270   &    19.62	(0.23)      &       19.79	(0.21)     &       19.00	(0.19)  &   0.17	(0.04)  &  0.10	(0.02) &   0.17	(0.03) \\
	3450    &    270   &    19.91	(0.34)      &       19.60	(0.19)     &       19.40	(0.26)  &   0.13	(0.04)  &  0.12	(0.02) &   0.12	(0.03) \\
	3720    &    270   &     -                  &       19.74   (0.23)     &       19.30    (0.26)  &    -              &  0.11	(0.02) &   0.13	(0.03) \\
	4005    &    270   &     -                  &       19.83   (0.28)     &        -               &    -              &  0.10 (0.03) &   - \\
    \end{tabular}
    \begin{tablenotes}
      \item Magnitudes are observed values, not corrected for Galactic extinction. Flux densities are corrected for Galactic and local extinction estimated in Sec. \ref{sec:SED}. Images are co-added with the given exposure time, except for images with 15 seconds exposure.
    \end{tablenotes}
\end{table*}
\begin{table}
	\centering
	\caption{Best-fitting temporal and spectral slopes of the GRB201015A light curve. The middle phase data is divided into two parts with a break time from 1635 to 1965 seconds. \label{tbl:1}}
	\begin{tabular}{c|c|c|c|c}
	\hline
	\hline
	Phases   & \multicolumn{3}{c}{$\alpha$ (error)}   &   $\beta$ (error)   \\
 & g\p        & r\p            & i\p                & 	        \\
\hline
Peak        &     0	    (0.10)  &  -0.11	(0.04)  &   -0.08	(0.05) &  -0.73	(0.08) \\
\hline
Post-peak   &  -1.53	(0.55)  &  -1.75	(0.26)  &   -1.11	(0.28) &  -0.48	(0.17) \\
\hline
\multirow{2}{4em}{Middle}&   -1.16	(0.44)  &  -1.06	(0.21)  &   -0.81	(0.24) &  -0.33	(0.19) \\
&   -0.86	(0.52)  &  -1.42	(0.27)  &   -1.37	(0.36) &  -0.01	(0.11) \\
    \end{tabular}
\end{table}

\begin{figure*}
	\includegraphics[width=\textwidth]{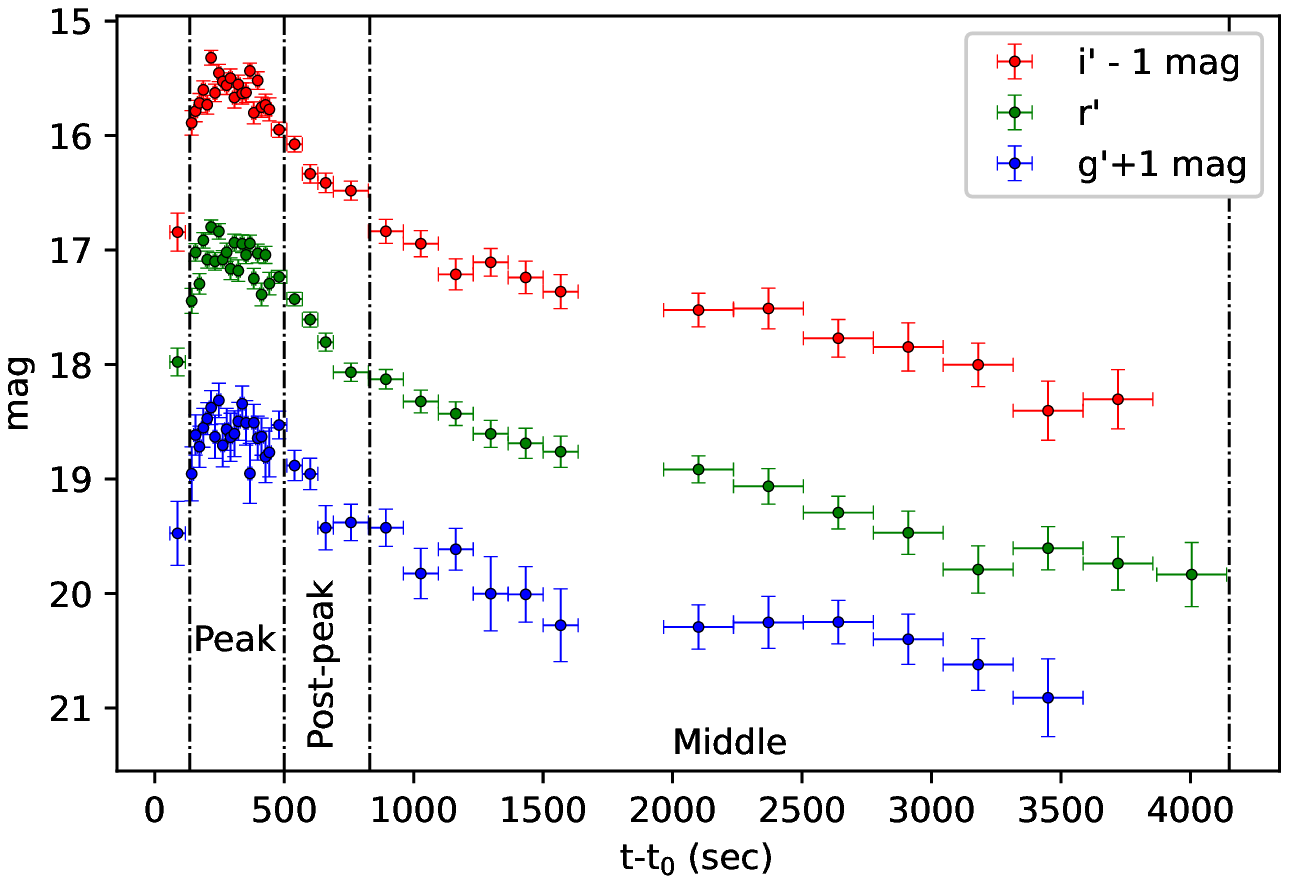}
	\caption{GRB201015A optical light curve. 
	The rise, during the first $\sim$ 250 seconds, is the expanding fireball which then cools.
	The magnitudes are not corrected for Galactic extinction.
    The i\p \ and g\p \ bands are artificially shifted for clarity -1 and +1 mag, respectively. 
        The vertical dash-dot lines separate the peak, post-peak, and middle phases. $\mathrm{t_0}$ is the time of the trigger. See text for explanation. 
		\label{fig:lightcurve}}
\end{figure*}

\section{Results} \label{results}
\subsection{Optical Light Curve}

The GRB201015A optical light curve (Fig.\ref{fig:lightcurve}) shows three main time periods of the early afterglow: the peak, post-peak and middle phases.
The  source rises to 17.3, 16.8 and 16.3 mag in g\p, r\p \ and i\p \ filters (without reddening corrections, see Table \ref{tbl:2}), respectively, during the first $\sim$ 250 seconds, and then fades as a power law ($t^{\alpha}$). The best-fitting temporal decay indices of the GRB 201015A optical light curve for the three time intervals are shown in Table \ref{tbl:1}. 

\cite{2020GCN.28653....1Z} reported a power law decay log slope of -1.2, between 78$<t<$1440 seconds, using Nanshan/NEXT r\p \ band observations. Our measured r\p \ band decay 
log slopes, -1.75$\pm$0.26 in the Post-peak phase and -1.06$\pm$0.21 in the middle phase, give
more detailed information in this time period.

\subsection{X-ray Light Curve}

\begin{figure*}
\includegraphics[width=\textwidth]{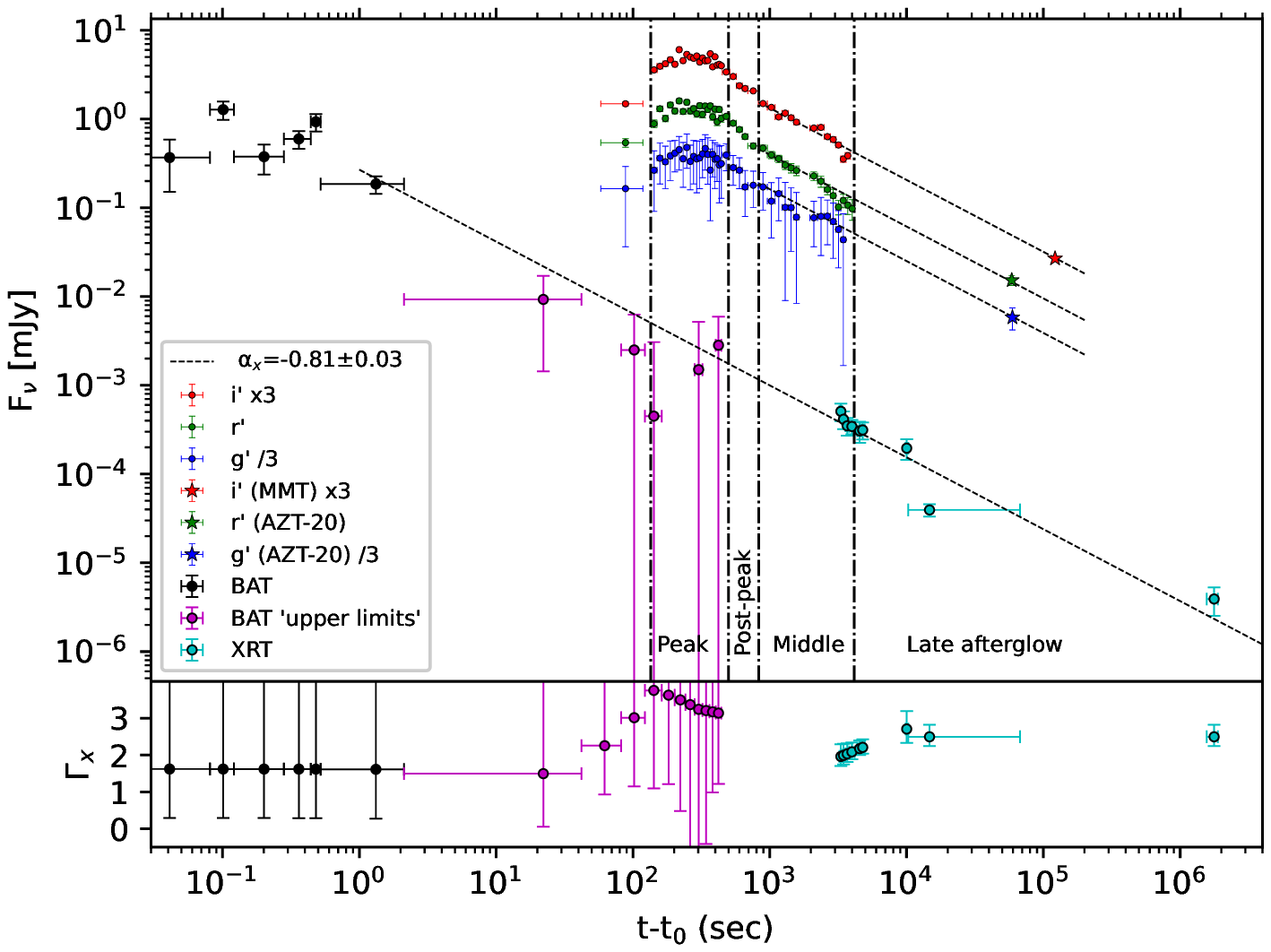}
	\caption{GRB201015A X-ray and optical light curves. The g',r',and i' light curves peak rapidly, then from a few thousand seconds, follow a power law decay with a log slope consistent with that measured in the X-rays (from a few s onward). The dashed lines indicate the X-ray temporal decay log slope $\alpha_X$=-0.81$\pm$ 0.03. 
		The 10 keV flux values of BAT and XRT were obtained from the UK \emph{Swift} Science Data Centre (for details, see text).  The upper error bars on $\Gamma_\mathrm{x}$ of BAT are extremely high ($\sim$17) and not shown on the plot.
		The optical light curves are corrected for both extinction within our Milky Way galaxy as well as local extinction within the host galaxy.
		The i\p \ and g\p \ bands are artificially shifted for clarity by multiplying by 3 and dividing by 3, respectively. 
		$\mathrm{t_0}$ is the time of the trigger.
  We refer to some BAT data points as, "BAT 'upper limits’" (those in purple color); though they are of low significance, they show the behavior of the X-ray emission between the prompt and peak phase is consistent with the indicated power law decay, and exclude any other dominant contributions during this time (see text). 
  The lower panel shows the photon index ($\Gamma_\mathrm{X}$) as a function of time.
		\label{fig:xray}}
\end{figure*}

The X-ray light curve is shown with the optical light curve in Fig.~\ref{fig:xray}.
We used the 10 keV flux values derived by the UK \emph{Swift} Science Data Centre at the University of Leicester from spectral and temporal fits of both the BAT (nominally 15-150 keV) and XRT (0.3-10 keV) measurements for a uniform comparison \citep[for details, see][]{2007A&A...469..379E,2009MNRAS.397.1177E,2010A&A...519A.102E}.
 We found an X-ray temporal decay log slope of $\alpha_X$=-0.81 $\pm$ 0.03 (dashed line in Fig. \ref{fig:xray}), which is the same as the Chandra X-ray decay slope reported by \cite{2020GCN.28822....1G}. 
To compare X-ray decay slope and the late time optical data from the MMT \citep[i\p \ band,][]{2020GCN.28676....1R} and AZT-20 \citep[r\p \ and g\p \ bands,][]{2020GCN.28673....1B} we plotted the $\alpha_X$ decay log slope over the g\p, r\p \ and i\p \, data, 
and we found the temporal decay log slope in optical and X-ray consistent at late times.

\subsection{Evolution of the spectral slope} 
\label{sec:SED}

In order to estimate the intrinsic log slope of the GRB emission, we had to correct for both extinction within our Milky Way galaxy as well as extinction closer to the burst. Galactic extinction in the direction of the GRB 201015A is 1.12, 0.77 and 0.58 mags for the Sloan g\p, r\p \  and i\p \, filters, respectively \citep[][]{2011ApJ...737..103S}. 
The local extinction can be estimated using the mean value $A_\mathrm{v}^\mathrm{local}$= 0.15 mags obtained by
\cite{2018ApJS..234...26L}, which was derived by ﬁtting a Gaussian to the distribution of published extinction values 
(log$_{10}$($A_\mathrm{v}^\mathrm{local}$) = -0.82$\pm$0.41).
We transformed the wavelengths of the
SED into 
the host galaxy frame 
using the redshift from \cite{2020GCN.28649....1D}.
Using the Small Magellanic Cloud (SMC) empirical extinction curve from \citet{1992ApJ...395..130P}\footnote{
In studies of active galactic nuclei and GRB only a small fraction of galaxies have a Milky Way like extinction curve \citep[see e.g.,][]{2011A&A...532A.143Z,1996ApJ...457..199G,2000PASP..112..537P}; we therefore assumed SMC-like extinction. \label{refnote}
}, we estimated the local extinction at Sloan g', r' and i' filters to be 0.19, 0.14 and 0.11
mags, respectively.

\begin{figure*}
    \begin{subfigure}{.45\textwidth}
    \centering
	\includegraphics[width=\textwidth]{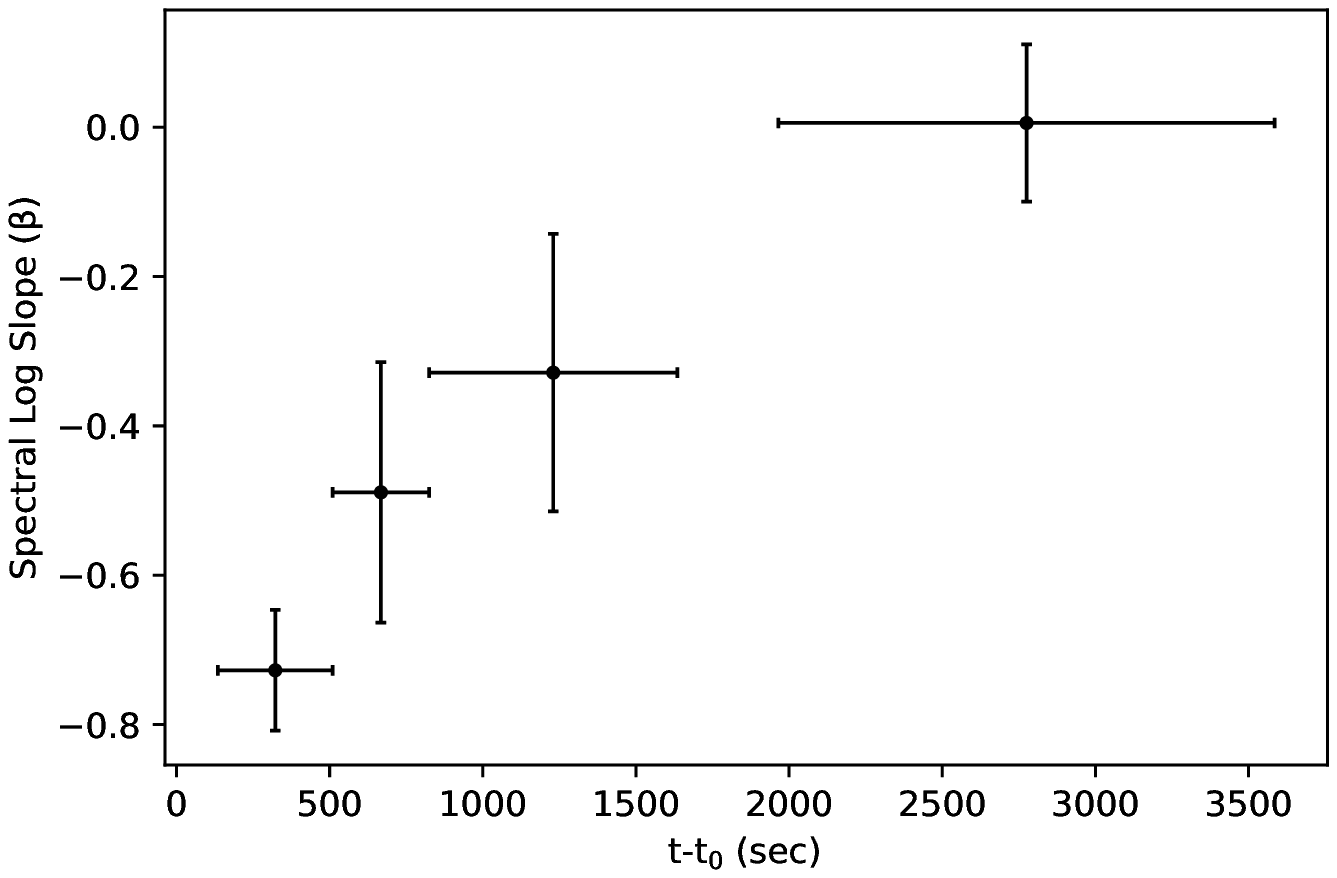}
	\caption{}
	\end{subfigure}
	\begin{subfigure}{.45\textwidth}
	\includegraphics[width=\textwidth]{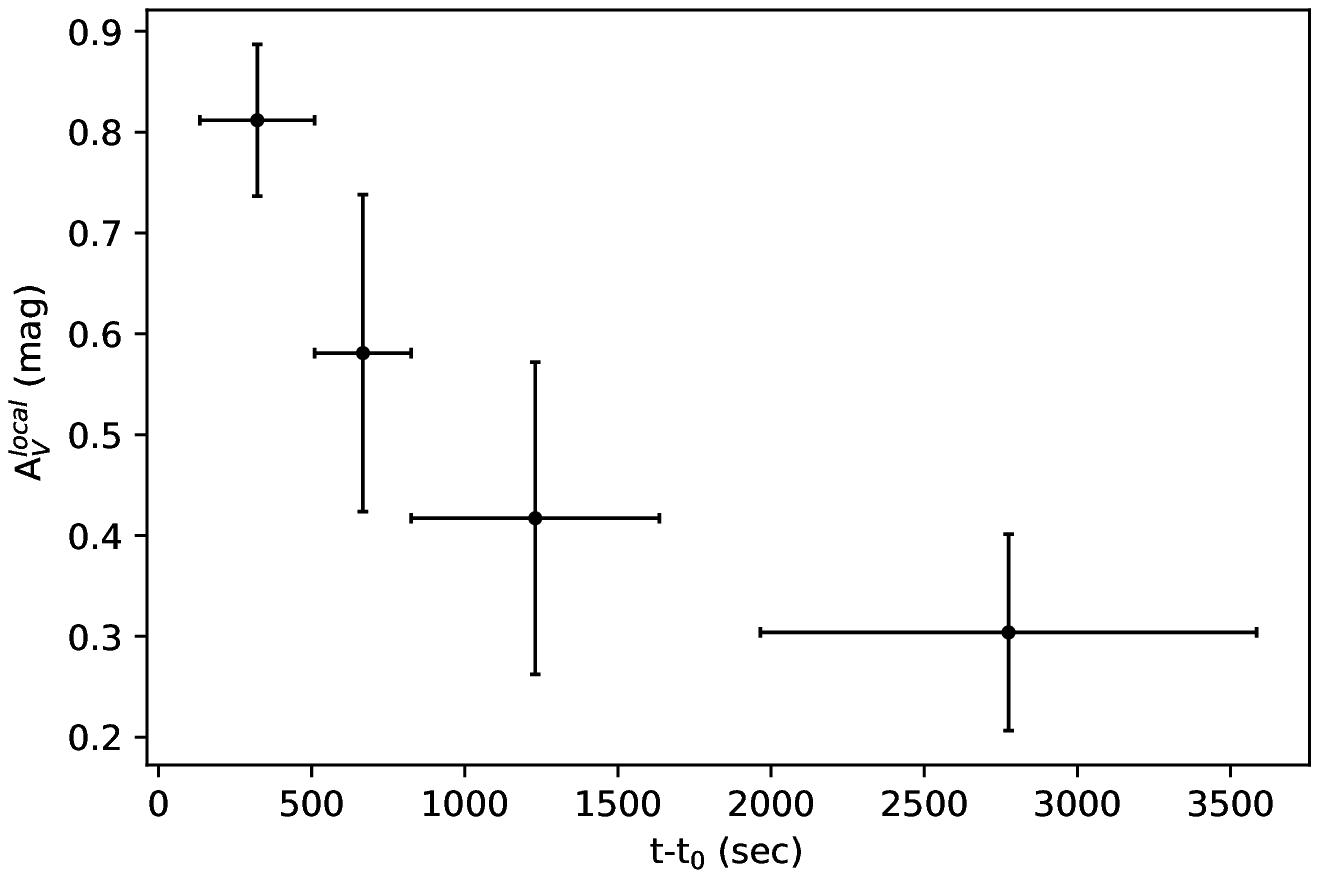}
	\caption{}
	\end{subfigure}
	\caption{(a) Evolution of the optical spectral shape. The log slope $\beta$ shows a large and rapid change from steep to flat (red to blue) in the first $\sim$ 2500 s. (b) Evolution of the extinction local to the source. Fixing the intrinsic optical spectral slope at the late-time value ($\beta$=0) and assuming a typical (SMC) extinction curve, we derive a rapid and substantial decrease in the extinction local to the source, $A_\mathrm{v}^\mathrm{local}$ . The $x$ axis error bars show the time intervals over which data were averaged. The $y$ axis error bars show the standard error of the mean value. $\mathrm{t_0}$ is the time of the trigger. \label{fig:beta}}
\end{figure*}

Figure \ref{fig:beta}(a) shows the optical spectral log slope as a function of time after applying the above corrections. 
The optical spectral log slope was determined by fitting a power law spectrum to the g\p, r\p \ and i\p \ data taken over time intervals from the peak (135 $\thicksim$ 510 seconds after the trigger), post-peak (510$\thicksim$825 seconds after the trigger) and middle phases (825$\thicksim$1635 seconds and 1968$\thicksim$3585 seconds after the trigger). The resulting spectral log slopes are -0.73$\pm$0.08, -0.49$\pm$0.17, -0.33$\pm$0.19 and 0.01$\pm$0.11, respectively. It is clear that the spectrum is getting bluer from the peak to the middle phase.

\subsection{Change in the local extinction} \label{sec:ext}	

If we assume that the change in spectral log slope is dominated by a decrease 
in the local extinction (such as would occur if dust was changed by the early bright GRB UV emission from the burst), we can estimate the local extinction as a function of time by assuming a constant intrinsic log slope given by that observed at late times, $\beta$=0. This assumption of a constant intrinsic optical spectral index is justified by the constant X-ray spectral index during this time,  \cite[i.e.  this interval is all during the the same "segment" of the burst in the terminology of][]{2006ApJ...642..354Z}, and a constant optical spectral index is observed in other GRBs (e.g., GRB 050315 and GRB050319). The assumption of 
 SMC-like dust is reasonable for extragalactic targets; though it is likely that departure from this extinction curve exists in detail \citep[e.g.,][]{2011AJ....141...36P}, and is significantly different in the UV, it is a reasonable approximation at optical wavelengths. We therefore derive $A_\mathrm{v}^\mathrm{local}$, the extinction as a function of time from both the host galaxy and that on smaller scales local to the burst, by applying the
 SMC extinction curve \citep[from][]{1992ApJ...395..130P} in the rest frame of the host. The result, shown in \ref{fig:beta}(b), shows the  $A_\mathrm{v}^\mathrm{local}$ decreasing from $~$0.8 mag to 0.3 mag in $\thicksim$2500 s between time bin centers.

\section{Discussion}
\label{discussion} 

The observed color variation in the optical at early time could arise from various scenarios:
\begin{enumerate}[ label=(\alph*)]
\item Two-component jet model: 
         \cite{2008Natur.455..183R} reported observations of the extraordinarily bright
optical and $\gamma$-ray emission of GRB 080319B that included prompt optical measurements starting $<$ 10 seconds after the $\gamma$-ray trigger and observations of the afterglow decay that continued for weeks. 
In order to explain the afterglow, which showed different decay indices ($\alpha$) at different times in X-rays and in the optical, they
proposed two jet components of different angular extent dominating emission at different times: an ultra-relativistic narrow jet, surrounded by a wide jet with a lower Lorentz factor. The evolving afterglow behavior is a result of both the forward and reverse external shocks from the narrow and wide jets dominating at different times. 
The paper shows that the X-ray observations are able to detect forward shock emission from both narrow and wide jets.

In our observations, the X-ray light curve decay of GRB 201015A shows only one power-law log slope for all of the afterglow time period, therefore a single component dominates emission, and the two-component jet scenario is not supported. 
\item Transition from reverse shock emission dominance to forward shock dominance:
\cite{2014Sci...343...38V} reported on the bright optical flash and fading afterglow from the powerful burst GRB 130427A, and showed a correlation of the optical and $>$ 100 MeV light curves during the first 7,000 seconds. Their simultaneous, multi-color, optical observations started at 132.9 seconds and continued until 7,585.9 seconds after trigger. At early times the observations are best explained by reverse shock emission generated in the relativistic burst ejecta as it collides with surrounding
material and at late times by a forward shock traversing the circumburst environment.
In our observations, the optical emission of the post-peak phase of GRB 201015A, between 500 $< t <$ 1600 seconds, may be interpreted as emission from the reverse shock associated with the interaction of the wide jet with   $\alpha_1\approx$-1.2 and $\beta_1$ = -0.42$\pm$0.14. 
The optical emission of the middle phase, between 1900 $< t <$ 4000 seconds, resembles late afterglow forward shock synchrotron emission from a decelerating relativistic shell that slows in an external medium with  $\alpha_2\approx$-0.8; $\beta_2$ = -0.01$\pm$0.11; and $\nu_\mathrm{m} < \nu_\mathrm{opt} < \nu_c$, where $\nu_m$, $\nu_\mathrm{opt}$ and $\nu_\mathrm{c}$ are the injection frequency, the optical band and the cooling frequency, respectively.
However, the colour change during these time periods, $\Delta\beta$ = 0.41, is much larger than seen in previous reverse/forward shock transitions \cite[e.g.,][]{2014Sci...343...38V}, and larger than expected by theory, $\Delta\beta\thicksim$0.25 \citep[e.g.,][]{1998ApJ...497L..17S}. 
Such a scenario for this burst is therefore inconsistent with theoretical predictions and previous examples of this kind of transition.

    \item Fading red prompt optical emission hypothesis: A bright red optical prompt emission component, if it were to fade at the time of the optical peak, would cause a red-to-blue color evolution, as was observed. We find this explanation problematic, however. In GRB 080319b, and similarly for other prompt optical light curves, a bright optical component, usually violently variable like the prompt gamma-ray emission, fades by orders of magnitude contemporaneously with the prompt gamma emission, and later the afterglow rises and dominates. For a burst where only the characteristic smooth rise of the afterglow is observed, especially in the case of an early rise, we acknowledge that it is difficult to strictly limit the contribution of any prompt component, e.g., to less than an order of magnitude weaker than the afterglow, at early times. However, we see a smooth rise to a peak, no violent variability in the early light curve, as is observed with strong prompt emission. The BAT T90 is less than 10 seconds for GRB 201015A, so we expect any significant prompt contributions to be negligible shortly afterward - far sooner than the  peak at $\sim$ 300 seconds. We see X-ray emission consistent with a classic power-law decay light curve from 1 second onward; indeed, by the time of peak BAT limits any prompt X-ray emission to less than a factor 30 fainter than during prompt emission. Any emission that scales with prompt X-rays must be very faint by this time. The spectral log slope in the optical is also inconsistent with that measured for the prompt X–gamma ray emission, so the optical cannot be the low-energy end of a single prompt power law spectrum. \citep[Theoretically speaking, we can add that in a general examination of prompt emission][found that in spectral cases besides the single power law extrapolation, prompt optical emission can be brighter (in F$_\mathrm{\nu}$) than at X as required, however, these cases have positive (very blue) optical log slopes, inconsistent with the requirement of a very red slope]{2009MNRAS.398.1936S}. We therefore reject the red prompt optical emission hypothesis for all the reasons above.

\item Time-varying dust extinction:
Long GRBs are associated with active star forming regions \citep{1993ApJ...405..273W,2006ARA&A..44..507W} and most GRBs show moderate dust extinction at $\thicksim$1000 seconds \citep{2009AJ....138.1690P}. \cite{2003ApJ...585..775P} suggested that dust destruction might cause a time-varying change in extinction A$_V$ (and possibly R$_V$) on relevant time scales,
and found that 
the extinction becomes less strong in the blue with time as a result of the faster sublimation of the smaller grains. 
\cite{2014MNRAS.440.1810M} have shown that time-varying dust extinction signatures could be contributing to the red-to-blue colour change in the first 200 s after the trigger of GRB 120119A. They found that the extinction A$_V$ is expected to decrease by an average of 0.61 ± 0.15 mag using best-fitting values.
We estimated the extinction in the first $\thicksim$1 hour (see Sec. \ref{sec:ext}, Fig. \ref{fig:beta}) to change from 0.81$\pm$0.08 mag at the peak phase to 0.3$\pm$0.1 mag at the last slope we measured (middle phase). 
This change $\Delta A_V$ of 0.51$\pm$0.13 is consistent with the value reported in \cite{2014MNRAS.440.1810M}, 
however, on a different time scale. Different dust columns or compositions could conceivably change this time scale.
The general consistency with both theoretical predictions and previous observations suggests that time-varying dust extinction correctly explains the gross features of our measured red-to-blue color evolution. We do not have sufficient spectral information to derive the actual extinction curve in detail, and deviations from our assumed typical SMC extinction seem likely at some level. The changing dust responsible for the changing extinction must be close to the burst, on a scale small compared to the size of the host galaxy. Sight lines to particular stars within our own Galaxy and others are known to have deviations from the average extinction curve, however, we argue that these effects are relatively small in the optical.  
For example, the extrema of the A$_{\mathrm\lambda}$/A$_\mathrm V$ curves presented in \cite{1989ApJ...345..245C} vary by less than 20\% down to 400 nm, the short wavelength limit of our g\p \ filter.
The reader is reminded that our quantitative results are dependent on the assumption of SMC-like dust, but only unusual dust would cause variations exceeding our uncertainties. 

\end{enumerate}

\section{Conclusion} \label{conclusion}

We have measured the flux of GRB 201015A in g\p, r\p, and i\p \ filter bands, and analysed the temporal and spectral properties of these data.
Our simultaneous multi-band observations of the early afterglow have shown the presence of a colour variation 
that cannot be explained by a two-component jet structure or a 
reverse/forward shock transition 
or a prompt emission component.

We find that the majority of the optical spectral slope $\beta$ evolution, coupled with no change in the X-ray temporal slope, are consistent with a monotonic decay of extinction. This provides evidence for dust destruction. 
Note that the same temporal slope for all bands suggests they are closely related (possibly co-located, at the leading edge of the forward shock).

This work shows that significant information about the early emission phase (and possibly prompt emission, if observed early enough) is being missed without such early observations with simultaneous multi-color instruments.
We plan to measure more GRBs, including their prompt emission, and create a catalog of spectral shape measurements covering as much of the wide variety of GRB behavior as possible.

\section*{Acknowledgements}
This research has been funded by the Science Committee of the Ministry of Science and Higher Education of the Republic of Kazakhstan (Grant Nos. AP14870504).
The NUTTelA-TAO Team acknowledges the support of the staff of the Assy-Turgen Astrophysical Observatory, Almaty, Kazakhstan, and the Fesenkov Astrophysical Institute, Almaty, Kazakhstan. Special thanks to ATO staff. This work made use of data supplied by the UK \emph{Swift} Science Data Centre at the University of Leicester, the MMT 6.5-meter telescope, and the AZT-20 telescope. EA is partially supported by Nazarbayev University Faculty Development Competitive Research Grant Program No 11022021FD2912.

\section*{Data Availability}

The data underlying this article are available in the article.








\bsp	
\label{lastpage}
\end{document}